\documentclass[runningheads]{llncs}
\usepackage{url}
\usepackage{blindtext, xcolor, graphicx, float, mathrsfs, amsmath, amsfonts}
\usepackage{booktabs}
\usepackage{multirow, minipage-marginpar}

\usepackage{threeparttable}
\begin{document}
\title{Semantic Feature Attention Network \\ for Liver Tumor Segmentation \\ in Large-scale CT database
}
%
%
\author{Yao Zhang\inst{1, 2}, Cheng Zhong\inst{3}, Yang Zhang\inst{3}, Zhongchao Shi\inst{3}, Zhiqiang He\inst{3}}
%
%
\institute{Institute of Computing Technology, Chinese Academy of Sciences \and University of Chinese Academy of Sciences \and Lenovo Research}
\maketitle              
\begin{abstract}
Liver tumor segmentation plays an important role in hepatocellular carcinoma diagnosis and surgical planning. In this paper, we propose a novel Semantic Feature Attention Network (SFAN) for liver tumor segmentation from Computed Tomography (CT) volumes, which exploits the impact of both low-level and high-level features. In the SFAN, a Semantic Attention Transmission (SAT) module is designed to select discriminative low-level localization details with the guidance of neighboring high-level semantic information. Furthermore, a Global Context Attention (GCA) module is proposed to effectively fuse the multi-level features with the guidance of global context. Our experiments are based on 2 challenging databases, the public Liver Tumor Segmentation (LiTS) Challenge database and a large-scale in-house clinical database with 912 CT volumes. Experimental results show that our proposed framework can not only achieve the state-of-the-art performance with the Dice per case on liver tumor segmentation in LiTS database, but also outperform some widely used segmentation algorithms in the large-scale clinical database.
\end{abstract}

\section{Introduction}

Liver cancer is one of the most common cancer diseases in the world and causes massive deaths every year~\cite{bray2018global}. The accurate measurements of liver tumor status from CT volumes, including tumor volume, shape and location, can assist doctors in making hepatocellular carcinoma evaluation and surgical planning. 

However, automatic liver tumor segmentation from the contrast-enhanced CT volumes is a very challenging task. First, the liver tumors have various sizes, shapes, textures and locations within the patients, therefore it is difficult to design features to extract the characteristics of liver tumors. Second, radiologists usually enhance CT volumes by an injection protocol for clearly observing tumors, which will introduce the noises inside the images. Meanwhile, different kinds of tumors (such as benign and malignant ones) have various appearances in different enhanced phases, which further poses challenges for automatic liver tumor segmentation. Third, there exists no clear boundaries for some liver tumors, which can bring the difficulties for both data annotation and segmentation.

Recent advances in deep learning is rapidly boosting performance in various medical applications~\cite{shen2017deep}~\cite{litjens2017a}. Several automatic liver tumor segmentation approaches based on fully convolutional network (FCN) have been proposed. This kind of models employs an encoder to extract and compress features from the input images and a decoder to restore the segmentation~\cite{lu2017automatic}. However, The ignorance of low-level features will hinder the generation of sharp prediction. Some methods thus employ skip connections between each level of the encoder and the decoder to recover the reduced spatial information caused by downsampling~\cite{Christ2016Automatic}~\cite{Sun2017Automatic}~\cite{li2018h}. Despite their success in preserving low-level features for more accurate segmentation, the characteristics and correlations of features from different levels of the encoder are ignored. Furthermore, most methods of this category intuitively fuse the features in a bottom-up way, i.e., from high level to low level, without the consideration of their diverse representations, which will lead to the problem of intra-class inconsistency.

To address these issues, we propose a novel SFAN for liver tumor segmentation from CT volumes. Attention mechanism is increasingly becoming a powerful tool to enhance the performance of deep neural networks and can focus on what we should put emphasis on. Inspired by~\cite{qin2018autofocus}, We design a SAT module to select the most effective features in low level. PSPNet~\cite{Zhao2017Pyramid} and Deeplab~\cite{chen2017rethinking} embed the context information from different scales to improve the consistency of network with the Pyramid Spatial Pooling module or Atrous Spatial Pyramid Pooling module. We draw their advantages and propose a GCA module to adaptively assign different weights to the features from different levels and effectively fuse them for more consistent and accurate segmentation. 

All previous liver tumor segmentation approaches are trained with limited amount of CT volumes. In contrast, our method leverages the knowledge of an annotated database of 912 CT volumes with various different scanning protocols (e.g., arterial and venous phase, and various resolution) and large variations in populations (e.g., ages and pathology). To the best of our knowledge, our experiment is the first time that nearly 1000 annotated CT volumes are adopted in liver tumor segmentation tasks. The experimental results demonstrate that the proposed SFAN outperforms some widely used segmentation algorithms in the large-scale clinical database. We also evaluate the SFAN on the public LiTS database and obtain the state-of-the-art performance.

Our main contributions can be summarized as follows:
\begin{itemize}
	\item We propose a SFAN to exploit the characteristics and correlation of low-level and high-level features.
	\item We design a SAT module that utilizes the neighboring high-level features to help select discriminative low-level features on each level in the transmission path between the encoder and the decoder.
	\item We present a GCA module that adaptively fuses multi-level features with the guidance of global context to improve the consistency of the segmentation.
	\item We evaluate the proposed SFAN on both the public LiTS database and the large-scale clinical database (912 CT volumes) and obtain the state-of-the-art performance.
\end{itemize}

The remainder of this paper is organized as follows. Section \ref{section:method} describes the proposed approach in details. Section \ref{section:experiment} shows and discusses the experiments and Section \ref{section:conclusion} draws the conclusion.


\begin{figure*}[!htp]
\centering
\includegraphics[width=\textwidth]{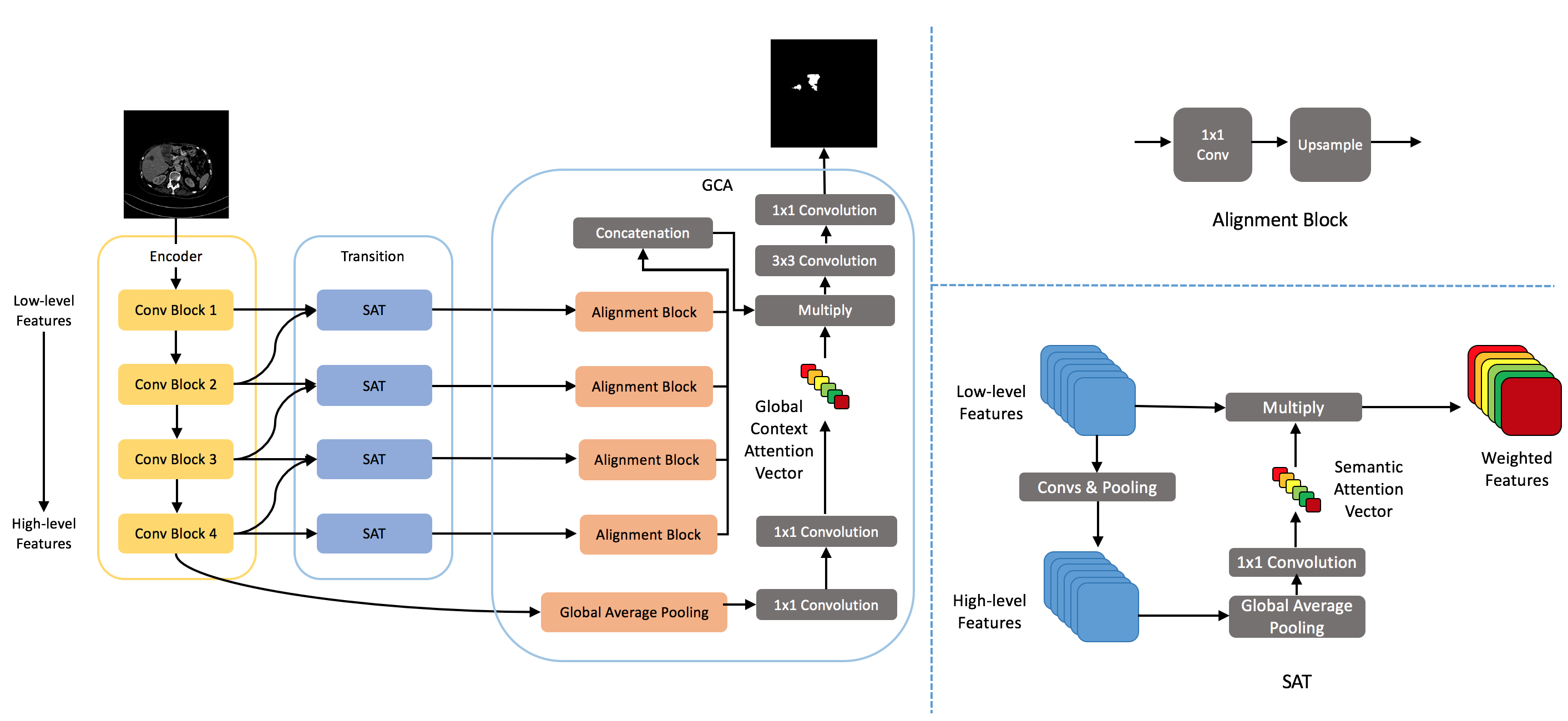}	
\caption{The architecture of the SFAN.}
\label{network}
\end{figure*}

\section{Method}
\label{section:method}
In this section, we first describe the complete architecture of the proposed SFAN composed of an encoder, a SAT module and a GCA module. Then we elaborate the design of SAT and GCA, and how these modules specifically handle the transmission and fusion of multi-level features. 

\subsubsection{Semantic Feature Attention Network}
We propose the SFAN for liver tumor segmentation based on an encoder-decoder architecture, as illustrated in Fig.~\ref{network}. We employ a convolutional neural network as the encoder to hierarchically extract different level features of the input CT image. Specifically, the encoder consists of 5 convolutional blocks and each followed by one max-pooling layer except for the last block. A convolution block is composed of 2 repeated convolution layers and each convolution layer is activated by a RELU. The filter size of all convolution layers is $3 \times 3$ and that of pooling layers is $2 \times 2$.
Instead of directly copying multi-level features from the encoder to the decoder, we design a SAT module to enhance the effectiveness of information transmission. According to the features of different levels, we propose a GCA module which to learn to weight the multi-level features by the global context and make the final semantic segmentation more accurate.

\subsubsection{Semantic Attention Transmission}
The input CT image can be extracted by the encoder into several levels according to the scale of the feature maps. In the lower level, the network encodes finer spatial textures. However, it has poor semantic consistency because of its small receptive field. While in the high level, it has strong semantic consistency due to large receptive view. However, only coarse prediction can be achieved because of the missing texture details. Therefore, to combine the advantages of both low-level and high-level features, we design a SAT module to weight the low-level features using the semantic information embedded in its neighboring high level features, which further enhances the feature transmission, as illustrated in Fig.~\ref{network}.

In details, high level features are compressed by 2 cascaded $1 \times 1$ convolution layers with sigmoid activation to generate a semantic attention vector. Then it is integrated with the low-level features by an element-wise multiplication. Then the weighted low-level features are transmitted to the decoder. If there is no higher level features (e.g., the last level of the network in Fig.~\ref{network}), the input features will directly go through this module without multiplied by a semantic attention vector.

Formally, $\bold{H}_{l}\in \mathbb{R}^{H\times W\times C}, l\in [0, L)$ denotes the feature maps produced by $l$-th level of the encoder, where $H$, $W$, $C$ represent the height, the width and the channel of the feature maps, and and $L$ denotes the number of levels respectively. We have the semantic attention vector $\bold{V_{t}}\in \mathbb{R}^{1\times 1\times C}$ as:
\begin{equation}
	\bold{V_{t}} = \sigma(Conv_{1\times 1}(Conv_{1\times 1}(f_g(\bold{H_{t+1}})))),
\end{equation}
Where $f_{g}(\cdot)$ means the global average pooling, $Conv(\cdot)$ denotes the convolution operation and $\sigma(\cdot)$ denotes the sigmoid activation. 
Then $\bold{H_{t}}$ are multiplied with $\bold{V_{t}}$ in an element-wise manner. The output semantic weighted low-level features $\bold{S_{t}}\in \mathbb{R}^{H\times W\times C}$ of the SAT module is formally given by $\bold{V_{t}}\bold{H_{t}}$.
As the high level features provide guidance information to low-level features to select the category localization details, SAT module makes the feature transmission more effective.

\subsubsection{Global Context Attention}
Unambiguously classifying tumors with different sizes in a CT scan requires different kinds of textures. For example, segmenting large tumors needs a large receptive view and global semantic information, while small tumors may require focusing on finer texture and local detailed information. Therefore, it is necessary to assign the high weights to the most discriminative and effective features according to the liver tumor properties. Motivated by this observation, we propose a GCA module, which employs the global average pooling to generate a global context attention vector to guide the fusion of the multi-level feature maps. The structure of GCA is depicted in Fig.~\ref{network}.

In details, the semantic weighted multi-level feature maps from the SAT module first go through an Alignment Block which can align the pyramid inputs for concatenation. The first component of this block is a $1\times 1$ convolutions that aligns the number of channels. And the followed upsample operation resizes all feature maps to the highest resolution of the inputs using bilinear interpolation. Finally the output feature maps of different ABs are concatenated to compose the multi-level feature maps.

The multi-level feature maps $\bold{P}$ $\in$ $\mathbb{R}^{H\times W\times (LC')}$ can be obtained as:

\begin{equation}
\bold{P} = [\bold{p}_{0}; \bold{p}_{1}; ...; \bold{p}_{l-1}], \ \bold{p}_{l} = f_{up}(Conv_{1\times 1}(\bold{S}_{l})),
\end{equation}
where $\bold{p}_{l}\in \mathbb{R}^{H\times W\times C'}$ denotes the aligned feature maps of $l$-th scale and $f_{up}(\cdot)$ represents the upsampling operation. 

The global attention branch consists of a global average pooling and 2 convolution layers. The global average pooling is in charge of compressing the feature maps of the input CT image to a global context vector. Then this vector goes through 2 cascaded $1\times 1$ convolution layers activated by a sigmoid function to transform the features along the channels and align the channels to the multi-level feature maps as an global context attention vector representing the different discrimination capabilities. Finally, the output of the GCA module is multiplied with global context attention vector and fed into a $3\times 3$ and a $1\times 1$ convolution to generate the segmentation probability map.

We can obtain the global context attention vector $\bold{G}\in \mathbb{R}^{H\times W\times (LC')}$ as:

\begin{equation}
\bold{G} = \sigma(Conv_{1\times 1}(Conv_{1\times 1}(f_{g}(\bold{H_{L-1}})))).
\end{equation}
The weighted multi-level features $\bold{O}$ can be given as $\bold{G}\bold{P}$.

\section{Experiments}
\label{section:experiment}
We quantitatively evaluate the proposed method on the public LiTS database and large-scale clinical database with a widely used metric Dice per case score~\cite{lits}. Since the CT volumes are from various sources and oriented differently, we normalize them by a rotation or/and a flip so that livers and backbones in all CT volumes are positioned at the left and bottom. Then we clip all CT volumes with a window $[-75, 175]$ HU (Hounsfield Unit) to remove the irrelevant background. A U-Net~\cite{ronneberger2015u} is trained to obtain a coarse segmentation of liver in advance. Then we truncate the liver region for the following liver tumor segmentation. 
We train all models from scratch on an NVIDIA Tesla P100 (with 16276M memory) GPU. The parameters are initialized with a Gaussian random initializer. Adam optimizer with an initial learning rate of $0.0001$ is used for parameters updating. During train stage, we employ weighted cross entropy~\cite{li2018h} as the loss function.

\begin{figure}[!tp]
\centering
\includegraphics[width=\linewidth]{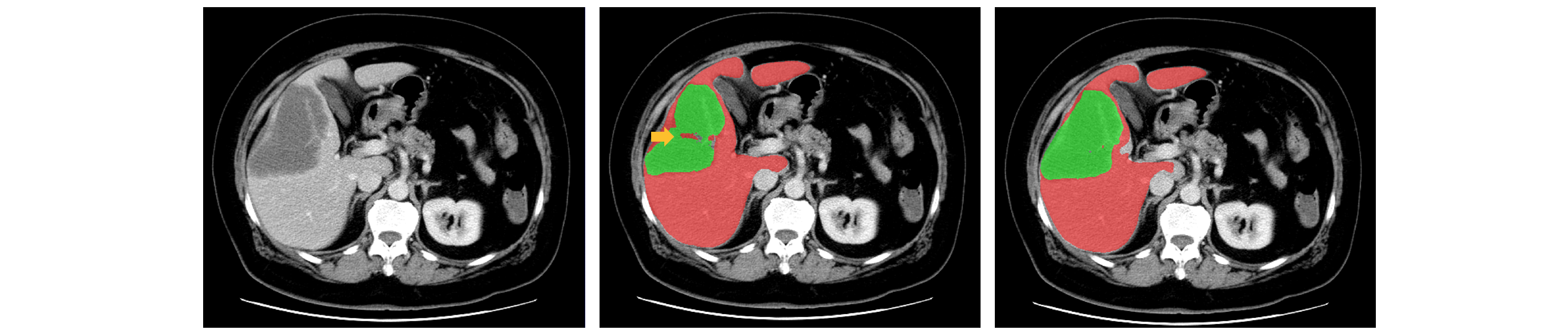}	
\caption{From left to right, it shows the CT image and the segmentation of U-Net and SFAN respectively. The inside holes (i.e., the intra-slice inconsistency) are highlighted by a yellow arrow.}
\label{lits_pred}
\end{figure}

\subsubsection{Experiments on the LiTS database}
The LiTS database~\cite{lits} consists of 131 and 70 contrast-enhanced abdominal CT volumes for training and testing. The CT volumes are acquired by different scanners and protocols from multiple clinical sites, with a largely varying XY spacing resolution from $0.55$ mm to $1.0$ mm and Z spacing resolution from $0.45$ mm to $6.0$ mm.

To further improve the accuracy of segmentation, we use a multi-scale inference (MI) strategy that takes image pyramids as inputs during the inference phase. Specifically, we resize a CT image to different resolutions to construct an image pyramid, and each of these CT images is fed into our model separately. Then all the resulting segmentation probability maps are resized to the original image size using bilinear interpolation. At last, these maps are merged to get the final prediction map using the average fusion strategy. Considering the tradeoff between accuracy and speed, here we use three scales as $\{0.5, 1.0, 1.5\}$.

\begin{figure}[!tp]
    \begin{minipage}[t]{0.45\linewidth}
    \centering
    \label{size}
    \includegraphics[width=2.25in]{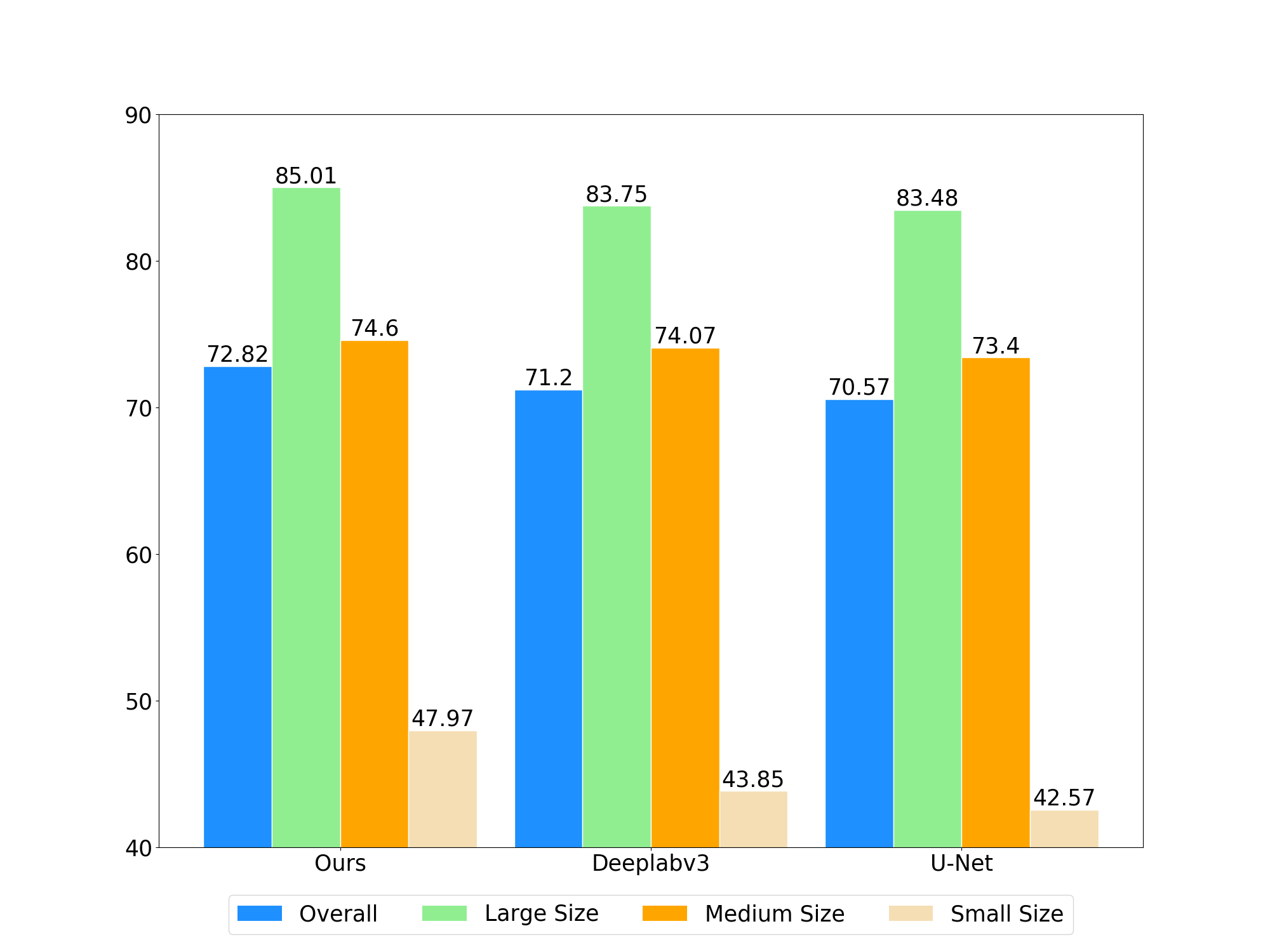}
    \caption{The results w.r.t. tumor size.}
    \end{minipage}
    \begin{minipage}[t]{0.47\linewidth}
    \label{phase}
    \centering
    \includegraphics[width=2.25in]{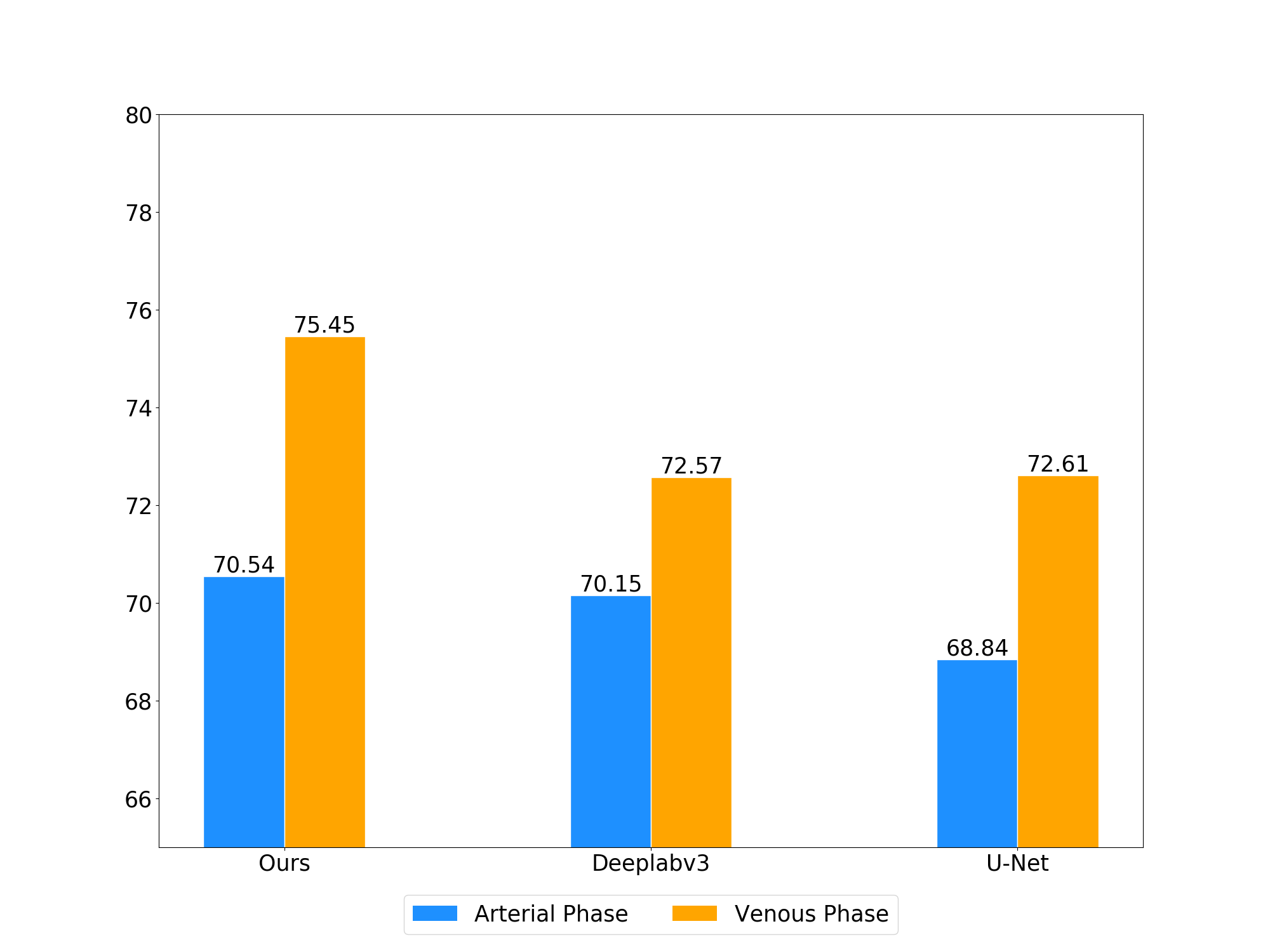}
    \caption{The results w.r.t. contrast phase.}
    \end{minipage}
\end{figure}

\subsubsection{Experiments on the Large-scale Clinical Database}
There are 912 contrast-enhanced CT volumes from 456 Chinese patients with both arterial and venous phases in the large-scale clinical database, which is collected from a top hospital. The properties of this database can be summarized as follows: (1) All the private information is removed; (2) To the best of our knowledge, this is the largest annotated liver tumor database of CT volumes, 4 times larger than LiTS database. All the annotations are verified by clinical doctors. (3) The database includes both venous and arterial phase of abdominal CT volumes. (4) All the CT volumes are in the resolution of $512\times 512$, with the XY spacing resolution ranges from $0.53\ mm$ to $0.91\ mm$ and Z spacing resolution ranges from $1.25\ mm$ to $1.5\ mm$. In our experiments, we randomly select 618 CT volumes for training and the remaining 294 CT volumes for testing. In addition, to precisely evaluate the segmentation performance related to the tumor staging, which is a clinical procedure aimed at documenting the anatomic extent of a malignant tumor, we further divide the testing database into 3 groups: small group including 120 cases with the size smaller than $5\ mm$, middle group including 110 cases with the size between $5\ mm$ and $10\ mm$ and large group including 64 cases with the size larger than $10\ mm$. The size of a tumor is represented by the max length of its pixel spacing along XYZ axes.

Our proposed method is compared with some widely used semantic segmentation algorithms, U-Net~\cite{ronneberger2015u} and DeepLabv3~\cite{chen2017rethinking}, which are implemented with open source code and kept the same experiment setting for fair comparison. With regard to DeepLabv3, we take ResNet-101 as the backbone and set output stride to 16. As for U-Net, we build a 5-level encoder and decoder, and set the number of features in first level to 64.

\subsubsection{Experimental Results and Discussion}

On the LiTS database, the proposed SFAN achieves 71.0\% on liver tumor segmentation in terms of Dice per case, which outperforms the 1st place method on MICCAI 2017 leaderboard. Fig.~\ref{lits_pred} presents an example of liver tumor segmentation results of the SFAN and U-Net on the LiTS database. We observe that the segmentation performance of SFAN can guarantee the intra-slice consistency. 
Although the best liver tumor segmentation performance is 73.8\% comparing with that of ours 71.0\% in LiTS open leaderboard, the advantages of SFAN can be summarized as follows: First our proposed method requires less computation cost and only needs 10 hours for training, which can avoid the hardware constraints for training complex models. In addition it is noteworthy that there are no 3D convolutions and no post-processing procedures adopted in our solution, which means there is great potentials for further improvement. Furthermore, SFAN is also validated on a large-scale clinical database with 912 CT volumes, 4 times larger than LiTS database, to demonstrate its effectiveness and generalization ability.

The comparison results on the large-scale clinical database are shown in Fig. 3 and Fig. 4. The proposed SFAN outperforms other segmentation algorithms in terms of the Dice per case, which illustrates the effectiveness and robustness of the proposed SFAN. From experimental results we can find that:
 \begin{itemize}
 \item Considering the tumor sizes, our proposed method achieves better performance than other segmentation algorithms in all 3 groups, as shown in Fig. 3. Notably, our proposed method has the largest improvement in the small tumor group with at least 4\% in terms of Dice per case. As we   know, if the tumor can be detected in its early stage, its prognosis will be better. Therefore, our proposed SFAN will be especially useful to find the small liver tumors to improve the diagnosis performance.
\item We also make some experiments considering the influence of different phases, as shown in Fig. 4. Compared with other widely used segmentation algorithms, our proposed method achieves better performance in both phases. We can also find that the segmentation performance in venous phase is obviously better than that in arterial phase because the tumor details in venous phase are more clear for most cases.
\end{itemize}

\section{Conclusion}
\label{section:conclusion}

In this paper, we have proposed a novel SFAN for liver tumor segmentation. In our method, first a SAT module is designed to embed the semantic information from high level to low-level features to enhance the feature transmission. Furthermore, a GCA module is proposed to effectively fuse the multi-level features using the global context to improve the consistency of the segmentation. Experimental results on the public LiTS demonstrate that our method achieves the state-of-the-art performance on liver tumor segmentation. We also evaluate the method on a large-scale clinical database with 912 CT volumes to demonstrate the effectiveness and robustness of the proposed SFAN.

%
%
%
\bibliographystyle{splncs04}
\bibliography{mybibliography}

\end{document}